\documentclass[]{IEEEphot}

\input ulem.sty

\jvol{}
\jnum{}
\jmonth{June}
\pubyear{2013}

\begin{document}

\title{Ultralow Phase Noise Microwave Generation from Mode-Locked Er-Fiber Lasers with Subfemtosecond Integrated Timing Jitter}

\author{Kwangyun Jung, Junho Shin, and Jungwon Kim~\IEEEmembership{Senior~Member,~IEEE}}

\affil{KAIST Institute for Optical Science and Technology and School of Mechanical, Aerospace and Systems Engineering, Korea Advanced Institute of Science and Technology (KAIST), Daejeon 305-701, South Korea}  


\maketitle

\markboth{}{Ultralow Phase Noise Microwave Generation from Mode-Locked Er-Fiber Lasers with Subfemtosecond Integrated Timing Jitter}


\begin{abstract}
We demonstrate ultralow phase noise 10-GHz microwave signal generation from a free-running mode-locked Er-fiber laser with -142 dBc/Hz and -157 dBc/Hz single-sideband absolute phase noise at 10 kHz and 100 kHz offset frequency, respectively. The absolute rms timing jitter is 1.5 fs when integrated from 1 kHz to 5 GHz (Nyquist frequency) offset frequency. In the 10 kHz - 10 MHz integration bandwidth typically used for microwave generators, the rms integrated jitter is 0.49 fs. The Er-fiber laser is operated in the stretched-pulse regime at close-to-zero dispersion to minimize the phase noise of extracted microwaves. In order to suppress the excess phase noise in the optical-to-electronic conversion process, we synchronize a low-noise voltage-controlled oscillator to the fiber laser using a fiber Sagnac-loop-based optical-microwave phase detector.
\end{abstract}


\section{Introduction}

Ultralow phase noise microwave and radio-frequency (RF) signal sources are important for various scientific and engineering applications. Some examples are driving accelerating cavities in free-electron lasers (FELs) \cite{Emma10}, ultrafast electron sources \cite{Gliserin12} and particle accelerators, high-performance antennas and radars, photonic analog-to-digital converters \cite{Valley07}, coherent communication, and high-speed, high-resolution signal analysis instrumentation, to name a few. With technical advances over more than a decade, commercially available sapphire-loaded cavity oscillators \cite{Locke08} and optoelectronic oscillators \cite{Yao96} can now generate extremely low phase noise microwave signals with $\sim$ -160 dBc/Hz level phase noise at 10 kHz offset frequency. Recently, optical frequency comb-based generation of ultralow phase noise microwave signals has attracted great interest. Many of them are based on the frequency division of the optical stability of ultrastable cavity-stabilized continuous-wave (CW) lasers via optical frequency combs (i.e., femtosecond mode-locked lasers) \cite{Fortier11,Millo09}. Although the optical frequency division can achieve microwave generation with extremely high phase stability and low phase noise, the installation and maintenance of ultrastable cavity-stabilized laser systems are technically difficult, and currently only a few advanced laboratories can fully utilize them. In addition, their usability outside well-controlled laboratory environment may be limited.

In fact, we have recently demonstrated that the high-frequency (e.g., $>$1 kHz offset frequency) timing jitter from free-running femtosecond mode-locked Er-fiber and Yb-fiber lasers can be engineered to be extremely low, well below 1 fs range \cite{TKKim11,Song11}. For many high-performance signal processing and communication applications  such as sampling clocks for analog-to-digital converters or master oscillators for signal source analyzers, the fast fluctuations in microwave phase and accumulated timing jitter impact the system performance most critically. Therefore, for applications that require ultralow short-term (e.g., $<$1 ms time scale or $>$1 kHz offset frequency) phase noise/timing jitter microwave signals, one may generate microwaves from a simple free-running mode-locked fiber laser generating optical pulse trains with ultralow intrinsic timing jitter. However, it has not been demonstrated that this sub-fs jitter level in the optical domain can be transferred to the equivalent phase noise in the microwave/RF domain. The achievable phase noise was limited by both the timing jitter performance of mode-locked laser itself and the excess phase noise in the optical-to-electronic conversion process.

In this paper, we investigate the achievable absolute phase noise of microwave signals generated from free-running mode-locked Er-fiber lasers. We synthesize 10-GHz microwave signals from 78-MHz free-running mode-locked Er-fiber lasers with -142 dBc/Hz and -157 dBc/Hz absolute single-sideband (SSB) phase noise at 10 kHz and 100 kHz offset frequency, respectively. The corresponding absolute rms timing jitter is 1.5 fs integrated from 1 kHz to 5 GHz (Nyquist frequency) offset frequency. Note that, in the typically used 10 kHz to 10 MHz integration bandwidth for microwave analog signal generators \cite{Agilent12}, the timing jitter is only 0.49 fs (rms). In order to mitigate the shot-noise-limited high-frequency phase noise floor in the optical-to-electronic conversion process (typically limited around -140 dBc/Hz level), we synchronize a low-noise voltage-controlled oscillator (VCO) to the mode-locked fiber laser using our recently demonstrated Fiber Loop-based Optical-Microwave Phase Detector (FLOM-PD) \cite{KJung12}. Note that in ref. \cite{KJung12}, it presented the method of synchronization (FLOM-PD) between a mode-locked laser and a microwave oscillator and showed that sub-femtosecond-level residual jitter and drift is possible when locking bandwidth is sufficient, regardless of the intrinsic phase noise of the sources in the bandwidth. In this paper, beyond having low residual noise in synchronization, we show that ultralow absolute phase noise is achievable in the microwave domain by combining  the subfemtosecond-jitter mode-locked fiber laser and the FLOM-PD-based synchronization method.  The demonstrated result shows that microwave/RF signals with sub-femtosecond-level absolute jitter are readily achievable from free-running mode-locked fiber lasers and commercially available VCOs, which enable more widespread use in various places, such as small-scale laboratories, large-scale scientific facilities (e.g., FELs and phased-array antennas), defense/radar systems and telecommunication stations, that require ultralow phase noise microwave/RF signals and/or optical pulse trains directly from relatively low-cost, compact and robust signal generators.

\section{Operation Principles}
Figure 1 shows the schematic for generating ultralow phase noise microwave signals from mode-locked fiber lasers. Due to the compatibility with standard low-cost, telecommunication-grade fiber components at 1550 nm, we use an Er-fiber laser as an optical master oscillator. In order to minimize the timing jitter of optical pulse trains from mode-locked fiber lasers, we operate the stretched-pulse Er-fiber lasers at the close-to-zero intra-cavity dispersion. This condition was theoretically predicted to have a minimal timing jitter \cite{Namiki97} and also was recently confirmed to reach sub-fs regime experimentally \cite{TKKim11,Song11}. In this way, in the low (below 40 kHz in this work) offset frequency range, the microwave phase noise is limited by the intrinsic timing jitter of mode-locked laser. In the higher offset frequency, the achievable phase noise is often limited by the optical-to-electronic conversion process \cite{Millo09,Quinlan11,Ivanov07}. When a standard $\sim$100-MHz repetition rate fiber laser is used, residual phase noise floor from direct photodetection is typically limited to $\sim$ -140 dBc/Hz level \cite{Quinlan11,JKim10}. In addition, amplitude-to-phase (AM-to-PM) conversion from direct photodetection can significantly degrade achievable microwave phase noise from mode-locked lasers in the lower offset frequency range as well \cite{Millo09,Ivanov07}. To overcome this limit down to $<$-160 dBc/Hz, several techniques such as repetition rate multiplication \cite{Haboucha11,Jiang11} have been recently demonstrated. In this work, instead of direct photodetection, we used the flywheel effect of a phase-locked loop (PLL) \cite{Fortier12,JKim04} based on a high-resolution FLOM-PD \cite{KJung12} and a VCO with low phase noise in the high ($>$1 MHz) offset frequency. The FLOM-PD is based on the electro-optic sampling by a quarter-wave-biased Sagnac-fiber-loop, and the phase error between the optical pulse train and the microwave signal applied to the phase modulator is converted to the intensity imbalance between the two output ports from the Sagnac loop. It has recently achieved long-term stable sub-fs-precision synchronization between an Er-fiber laser and a microwave VCO with a low AM-to-PM conversion coefficient (0.06 rad/($\triangle$P/P0)). More detailed information on the design and implementation of the FLOM-PD can be found in ref. 11. By lowering the residual phase noise in the FLOM-PD and using a VCO with good phase noise performance in high offset frequency, one can mitigate both the shot noise floor and the AM-to-PM conversion problems in the optical-to-electronic conversion process, as illustrated in Fig. 1.

\begin{figure}[t]
\centering
\includegraphics[width=30pc]{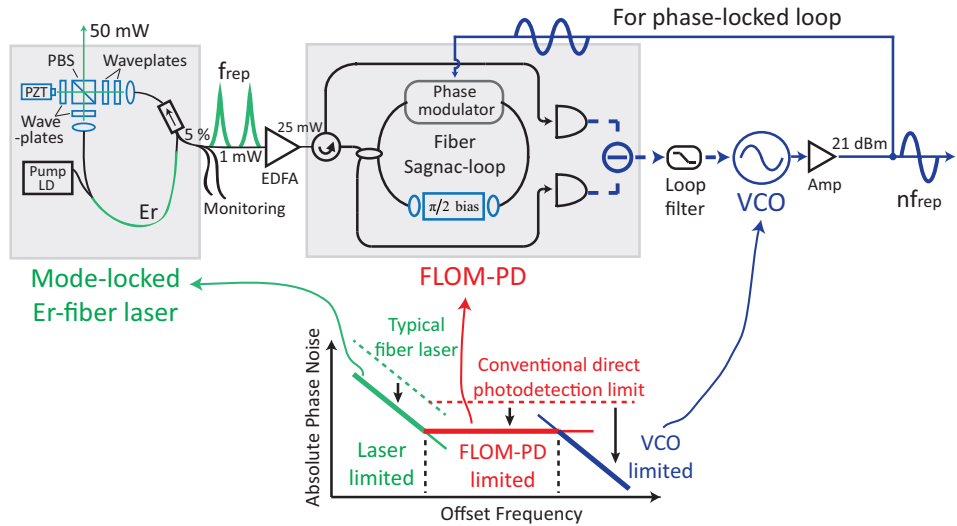}
\caption{Principle of ultralow phase noise microwave generation from a free-running mode-locked fiber laser and a VCO phase-locked by a fiber loop optical-microwave phase detector (FLOM-PD). The combination of (a) the optimized timing jitter performance in the mode-locked fiber laser, (b) the use of FLOM-PD with -160 dBc/Hz level residual phase noise, and (c) the use of low phase noise VCO enables the ultralow phase noise microwave generation.}
\label{fig_env1}
\end{figure}

\section{Experimental Setup}

Figure 2 shows the schematic of the experimental setup. Two 78-MHz repetition rate, nonlinear polarization rotation (NPR)-mode-locked, stretched-pulse Er-fiber lasers operating at close-to-zero net cavity dispersion \cite{TKKim11} are built and used as optical master oscillators for microwave generation. From the hybrid coupler isolator inside the laser cavity (where the coupling port is located in front of the isolator in a reflective manner, see the isolator symbol in Fig. 1), 6 mW (5 \% of intra-cavity power) is extracted and used for the microwave generation, repetition rates locking between two lasers, and laser condition monitoring. For the absolute phase noise measurement of the synthesized microwave signals, we built two almost identical fiber lasers and microwave extraction systems. Repetition rates of the two independent lasers were locked to each other with $<$400 Hz bandwidth using conventional direct photodetection and a frequency mixer. Therefore, we can measure the absolute phase noise of the microwave signals above $\sim$1 kHz offset frequency. After amplified to 25 mW by the Er-doped fiber amplifier, the optical pulse train is applied to the FLOM-PD. In this work, instead of long-term stable synchronization purpose shown in ref. 11, we used the FLOM-PD to reduce the absolute phase noise level of the generated microwave signals to the -160 dBc/Hz level. A 10-GHz dielectric resonator oscillator (DRO), manufactured by INWAVE AG, is used as the VCO in this work. The 10-GHz, +12 dBm microwave signal (the 129th harmonic of the pulse repetition rate) from the VCO is amplified by a low-noise amplifier (Nextec-RF, NB00424) to the +21 dBm. After split by a 3-dB splitter, +18 dBm of 10-GHz microwave is applied to the optical-microwave phase detector, and the other +18 dBm is applied to the out-of-loop interferometric microwave phase noise measurement setup. By closing the PLL, the 10-GHz VCO is phase-locked to the 78-MHz Er-fiber laser. To resolve ultralow phase noise of the VCO at high offset frequency, we built and used an interferometric measurement setup using microwave carrier suppression \cite{Ivanov98}. Note that the locking bandwidth between the two lasers ($\sim$400 Hz) is set to the minimal bandwidth that allows enough carrier suppression in the interferometric microwave phase measurement. 

\begin{figure}[t]
\centering%
\includegraphics[width=30pc]{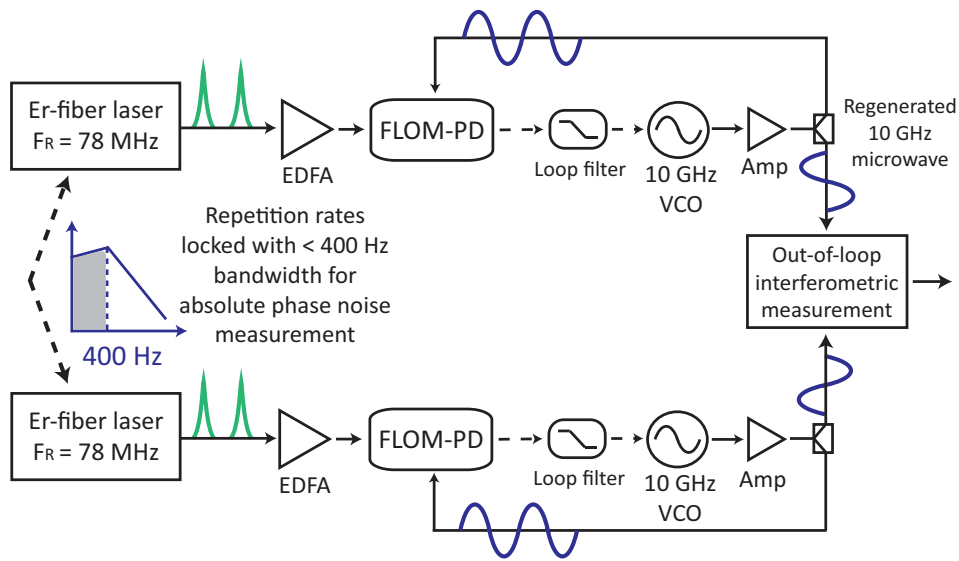}
\caption{Experimental setup for microwave signal generation and characterization}
\label{fig_env2}\vspace*{-6pt}
\end{figure}

\section{Results}

Figure 3 shows the measurement result of a single-sideband (SSB) phase noise at 10 GHz carrier frequency. Curve (i) is the absolute phase noise of the phase-locked VCO to the mode-locked Er-fiber laser. In order to evaluate the phase noise of a single system, the measured phase noise was subtracted by 3 dB assuming the equal phase noise contributions from two almost identical lasers and VCOs. In the 10 kHz $-$ 40 kHz offset frequency range, this curve follows the -20 dB/decade slope, which indicates the random walk nature of the quantum-limited phase noise in free-running mode-locked laser. At offset frequency lower than 10 kHz, the slope slightly increases to $\sim$ -25 dB/decade, which seems to be caused by technical noise (such as acoustic noise) and/or intensity-noise coupled jitter (such as by self-steepening effect) in the mode-locked fiber lasers. At 50 kHz, it reaches the phase detector resolution limit at $\sim$ -157 dBc/Hz (curve (ii)) caused by the phase detector background noise. Outside the locking bandwidth (at $\sim$1 MHz), this phase noise follows the free-running VCO phase noise (curve (iii)). As a result, we can take advantage of low phase noise from the VCO in the high offset frequency where the phase noise of mode-locked lasers cannot be transferred in the optical-to-electronic conversion process. The integrated rms absolute timing jitter of the synthesized microwave signal is 1.5 fs when integrated from 1 kHz to 5 GHz (Nyquist frequency) offset frequency, and the white phase noise floor is below -180 dBc/Hz. Note that the -185 dBc/Hz noise floor above 100 MHz offset frequency already contributes $>$1 fs integrated jitter. For the 10 kHz $-$ 10 MHz offset frequency ranges typically used for quoting timing jitter in microwave generators for telecommunication and signal processing applications \cite{Agilent12}, the integrated jitter is only 0.49 fs (rms). This sub-fs jitter level in microwaves is possible mainly due to the sub-fs-level jitter of optical pulse trains from mode-locked lasers, as the integrated jitter from free-running VCO is $\sim$60 fs from 1 kHz to 5 GHz and the absolute phase noise is suppressed by $>$30 dB by the laser in the $<$100 kHz offset frequency range, where the integrated jitter is dominated.

\begin{figure}[t]
\centering%
\includegraphics[width=30pc]{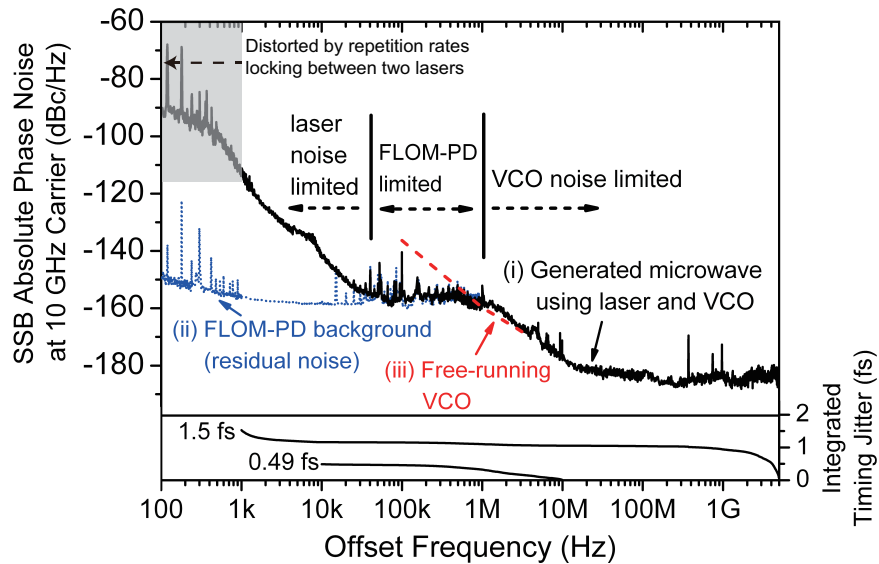}
\caption{Absolute phase noise measurement results. Curve (i), Phase noise of the generated microwaves from mode-locked fiber lasers and locked VCOs, measured by an interferometric microwave phase noise measurement technique. The absolute rms timing jitter integrated from 1 kHz to 5 GHz (from 10 kHz to 10 MHz) offset frequency is 1.5 fs (0.49 fs). Curve (ii), Fiber-loop optical-microwave phase detector (FLOM-PD) residual phase noise floor. Curve (iii), Phase noise of a free-running VCO used.}
\label{fig_env3}\vspace*{-6pt}
\end{figure}

Note that the phase noise increases with -20 dB/dec or higher in the lower offset frequency because the mode-locked laser is a free-running oscillator. If necessary, one can lock the free-running fiber laser to the commercial high-performance microwave sources (such as \cite{Agilent12}) or benchtop Rb frequency standards for better long-term phase stability in the $<$1 kHz range. Also note that we recently demonstrated that the timing jitter of Yb-fiber lasers can be reduced down to the 20-attosecond level [10 kHz $-$ 94 MHz] by reducing the non-gain fiber length \cite{HKim13} and a similar fiber laser design may be applicable to Er-fiber lasers in the near future, which will enable further phase noise reduction in microwave signals extracted from the Er-fiber lasers. If a similar performance laser is used, the projected achievable phase noise is -160 dBc/Hz at 10 kHz offset frequency for 10-GHz microwaves. Although we used a FLOM-PD for robust, long-term stable phase-locking between the laser and the VCO with low AM-to-PM conversion coefficient, we believe that one can also use different schemes of phase-locking as well, for example, based on repetition rate multiplication \cite{Haboucha11,Jiang11} and advanced direct photodetection techniques \cite{Taylor11,Zhang12} with lower residual phase noise floor. By using fs-precision timing-stabilized fiber links \cite{JKim08}, it will be also possible to regenerate sub-fs absolute jitter microwave/RF signals at multiple remote locations that are phase-locked to the optical pulse trains from femtosecond mode-locked lasers, which will be particularly useful in the ultrafast electron sources, FELs, particle accelerators, or distributed antennas/radars applications.

\section{Summary}

 For the first time to our knowledge, we have generated 10-GHz microwave signals from free-running femtosecond mode-locked lasers with $<$-140 dBc/Hz phase noise level at 10 kHz offset frequency and 1.5 fs [1 kHz $-$ 5 GHz] (0.49 fs [10 kHz $-$ 10 MHz]) integrated rms timing jitter. We showed that, in the high offset frequency range (e.g., $>$1 kHz), well-designed free-running mode-locked fiber lasers, assisted by an effective optical-to-microwave conversion method, can achieve absolute phase noise performance that was only possible by cavity-stabilized sources previously. We are currently also exploring the possibility of an additional fiber-based reference to further improve the phase noise in the lower offset frequency range. Our approach has a unique advantage that it can generate ultralow timing jitter optical pulse train and ultralow phase noise microwaves simultaneously from a rack-sized instrument. In addition, they are phase-locked to each other. This property is useful for many applications such as photonic ADCs, optical sampling, timing and synchronization of FELs and ultrafast electron sources, where both low-noise optical pulse trains and microwaves are necessary. 

\section*{Acknowledgements}
This research was supported by the National Research Foundation (NRF) of South Korea \\(2012R1A2A2A01005544).

\bibliographystyle{IEEEtran}
\bibliography{thesis}

\begin{thebibliography}{24}
\providecommand{\url}[1]{#1}
\csname url@samestyle\endcsname
\providecommand{\newblock}{\relax}
\providecommand{\bibinfo}[2]{#2}
\providecommand{\BIBentrySTDinterwordspacing}{\spaceskip=0pt\relax}
\providecommand{\BIBentryALTinterwordstretchfactor}{4}
\providecommand{\BIBentryALTinterwordspacing}{\spaceskip=\fontdimen2\font plus
\BIBentryALTinterwordstretchfactor\fontdimen3\font minus
  \fontdimen4\font\relax}
\providecommand{\BIBforeignlanguage}[2]{{%
\expandafter\ifx\csname l@#1\endcsname\relax
\typeout{** WARNING: IEEEtran.bst: No hyphenation pattern has been}%
\typeout{** loaded for the language `#1'. Using the pattern for}%
\typeout{** the default language instead.}%
\else
\language=\csname l@#1\endcsname
\fi
#2}}
\providecommand{\BIBdecl}{\relax}
\BIBdecl

\bibitem{Emma10}
P. Emma, R. Akre, J. Arthur, R. Bionta, C. Bostedt, J. Bozek, A. Brachmann, P. Bucksbaum, R. Coffee, F.-J. Decker, Y. Ding, D. Dowell, S. Edstrom, A. Fisher, J. Frisch, S. Gilevich, J. Hastings, G. Hays, Ph. Hering, Z. Huang, R. Iverson, H. Loos, M. Messerschmidt, A. Miahnahri, S. Moeller, H.-D. Nuhn, G. Pile, D. Ratner, J. Rzepiela, D. Schultz, T. Smith, P. Stefan, H. Tompkins, J. Turner, J. Welch, W. White, J. Wu, G. Yocky, and J. Galayda, ''First lasing and operation of an angstrom-wavelength free-electron laser," Nat. Photon., vol. 4, no. 9, pp. 641-647, Sep. 2010.

\bibitem{Gliserin12}
A. Gliserin, A. Apolonski, F. Krausz, and P. Baum, "Compression of single-electron pulses with a microwave cavity," New J. Phys., vol. 14, no. 7, pp. 073055-1$-$073055-18, Jul. 2012.

\bibitem{Valley07}
G. C. Valley, "Photonic analog-to-digital converters," Opt. Exp., vol. 15, no. 5, pp. 1955-1982, Mar. 2007.

\bibitem{Locke08}
C. R. Locke, E. N. Ivanov, J. G. Hartnett, P. L. Stanwix, and M. E. Tobar, "Design techniques and noise properties of ultrastable cryogenically cooled sapphire-dielectric resonator oscillators," Rev. Sci. Instrum., vol. 79, no. 5, pp. 051301-1$-$051301-12, May. 2008.

\bibitem{Yao96}
X. S. Yao and L. Maleki, "Optoelectronic microwave oscillator," J. Opt. Soc. Am. B, vol. 13, no. 8, pp. 1725-1735, Aug. 1996.

\bibitem{Fortier11}
T. M. Fortier, M. S. Kirchner, F. Quinlan, J. Taylor, J. C. Bergquist, T. Rosenband, N. Lemke, A. Ludlow, Y. Jiang, C. W. Oates, and S. A. Diddams, "Generation of ultrastable microwaves via optical frequency division," Nat. Photon., vol. 5, no. 7, pp. 425-429, Jul. 2011.

\bibitem{Millo09}
J. Millo, M. Abgrall, M. Lours, E. M. L. English, H. Jiang, J. Gu\'ena, A. Clairon, M. E. Tobar, S. Bize, Y. Le Coq, and G. Santarelli, "Ultralow noise microwave generation with fiber-based optical frequency comb and application to atomic fountain clock," Appl. Phys. Lett., vol. 94, no. 14, pp. 141105-1$-$141105-3, Apr. 2009.

\bibitem{TKKim11}
T. K. Kim, Y. Song, K. Jung, C. Kim, H. Kim, C. H. Nam, and J. Kim, "Sub-100-as timing jitter optical pulse trains from mode-locked Er-fiber lasers," Opt. Lett., vol. 36, no. 22, pp. 4443-4445, Nov. 2011.

\bibitem{Song11}
Y. Song, C. Kim, K. Jung, H. Kim, and J. Kim, "Timing jitter optimization of mode-locked Yb-fiber lasers toward the attosecond regime," Opt. Exp., vol. 19, no. 15, pp. 14518-14525, Jul. 2011.

\bibitem{Agilent12}
Agilent Technologies, Agilent E8257D PSG Microwave Analog Signal Generator Data Sheet, Literature number 5989-0698EN (2012).

\bibitem{KJung12}
K. Jung and J. Kim, "Subfemtosecond synchronization of microwave oscillators with mode-locked Er-fiber lasers," Opt. Lett., vol. 37, no. 14, pp. 2958-2960, Jul. 2012.

\bibitem{Namiki97}
S. Namiki and H. A. Haus, "Noise of the Stretched Pulse Fiber Laser: Part I - Theory," IEEE J. Quantum Electron., vol. 33, no. 5, pp. 649-659, May. 1997.

\bibitem{Quinlan11}
F. Quinlan, T. M. Fortier, M. S. Kirchner, J. A. Taylor, M. J. Thorpe, N. Lemke, A. D. Ludlow, Y. Jiang, and S. A. Diddams, "Ultralow phase noise microwave generation with an Er:fiber-based optical frequency divider," Opt. Lett., vol. 36, no. 16, pp. 3260-3262, Aug. 2011.

\bibitem{Ivanov07}
E. N. Ivanov, J. J. McFerran, S. A. Diddams, and L. Hollberg, "Noise properties of microwave signals synthesized with Femtosecond Lasers," IEEE Trans. Ultrason. Ferroelectr. Freq. Control, vol. 54, no. 4, pp. 736-745, Apr. 2007.

\bibitem{JKim10}
J. Kim and F. X. K\"artner, "Microwave signal extraction from mode-locked lasers with attosecond relative timing drift," Opt. Lett., vol. 35, no. 12, pp. 2022-2024, Jun. 2010.

\bibitem{Haboucha11}
A. Haboucha, W. Zhang, T. Li, M. Lours, A. N. Luiten, Y. Le Coq, and G. Santarelli, "Optical-fiber pulse rate multiplier for ultralow phase-noise signal generation," Opt. Lett., vol. 36, no. 18, pp. 3654-3656, Sep. 2011.

\bibitem{Jiang11}
H. Jiang, J. Taylor, F. Quinlan, T. Fortier, and S. A. Diddams, "Noise Floor Reduction of an Er:Fiber Laser-Based Photonic Microwave Generator," IEEE Photon. J., vol. 3, no. 6, pp. 1004-1012, Dec. 2011.

\bibitem{Fortier12}
T. M. Fortier, C. W. Nelson, A. Hati, F. Quinlan, J. Taylor, H. Jiang, C. W. Chou, T. Rosenband, N. Lemke, A. Ludlow, D. Howe, C. W. Oates, and S. A. Diddams, "Subfemtosecond absolute timing jitter with a 10 GHz hybrid photonic-microwave oscillator," Appl. Phys. Lett., vol. 100, no. 23 , pp. 231111-1$-$231111-3, Jun. 2012.

\bibitem{JKim04}
J. Kim, F. X. K\"artner, and M. H. Perrott, "Femtosecond synchronization of radio frequency signals with optical pulse trains," Opt. Lett., vol. 29, no. 17, pp. 2076-2078, Sep. 2004.

\bibitem{Ivanov98}
E. N. Ivanov, M. E. Tobar, and R. A. Woode, "Microwave Interferometry: Application to Precision Measurements and Noise Reduction Techniques," IEEE Trans. Ultrason. Ferroelectr. Freq. Control, vol. 45, no. 6, pp. 1526-1536, Nov. 1998.

\bibitem{HKim13}
H. Kim, P. Qin, J. Shin, Y. Song, C. Kim, K. Jung, C. Wang, and J. Kim, "Reduction of timing jitter to the sub-20-attosecond regime in free-running femtosecond mode-locked fiber lasers," Paper CTh4M.4, to be presented at Conference on Lasers and Electro Optics (CLEO), San Jose, USA, 9-14 June 2013.

\bibitem{Taylor11}
J. Taylor, S. Datta, A. Hati, C. Nelson, F. Quinlan, A. Joshi, and S. Diddams, "Characterization of Power-to-Phase Conversion in High-Speed P-I-N Photodiodes," IEEE Photon. J., vol. 3, no. 1, pp. 140-151, Feb. 2011.

\bibitem{Zhang12}
W. Zhang, T. Li, M. Lours, S. Seidelin, G. Santarelli, Y. Le Coq, "Amplitude to phase conversion of InGaAs pin photodiodes for femtosecond lasers microwave signal generation," Appl. Phys. B, vol. 106, no. 2, pp. 301-308, Feb. 2012.

\bibitem{JKim08}
J. Kim, J. Cox, J. Chen, and F. X. K\"artner, "Drift-free femtosecond timing synchronization of remote optical and microwave sources," Nat. Photon., vol. 2, no. 12, pp. 733-736, Dec. 2008.

\end{thebibliography}



\end{document}